# Range restriction, admissions criteria, and correlation studies of standardized tests

Alexander Small[1]

Department of Physics and Astronomy

California State Polytechnic University, Pomona

**Abstract:**

Recent and influential critiques of standardized testing have noted the existence of non-trivial numbers of successful scientists who received low scores on the GRE. Here we use computer simulations to show that the prevalence of such examples is consistent with the reasonable hypothesis that academic performance depends on multiple variables. We examine the effects of admissions criteria on the observed predictive power of different variables, and show that observed correlations between student performance and student characteristics depend as much on the method of selecting students as on causal relationships detectable in the wider applicant pool. This is an example of the well-known statistical phenomenon of range restriction, and we offer relevant caveats and recommendations for further studies of admissions tests. We also show that the magnitude of range restriction effects depend on how student characteristics are weighted in admissions decisions.

[1] Department of Physics and Astronomy

California State Polytechnic University

3801 West Temple Avenue

Pomona, CA 91768

arsmall@cpp.edu

909-869-5202

## Introduction

The Graduate Record Examinations (GREs) are subjects of longstanding controversy, especially because of effects on groups that are under-represented in graduate education (W. E. Sedlacek, 2004). Recently the issue has received particular attention in the physical sciences (Miller, 2013; Miller & Stassun, 2014; W. Sedlacek, 2014).  An influential recent study examined the Physics GRE Subject Test scores of astronomers who secured competitive postdoctoral fellowships, and found wide variability in their GRE scores (Levesque, Bezanson, & Tremblay, 2015). The authors concluded that "we find no evidence that the PGRE [Physics GRE Subject Test] can be used as an effective predictor of 'success' either in or beyond graduate school." Other astronomers have expressed well-motivated concern about "a deep-seated and unfounded belief that these test scores are good measures of ability, of potential for doing well in graduate school and of long-term potential as a scientist" (Miller & Stassun, 2014), citing examples of programs that have had success in training students with low GRE scores (Stassun et al., 2011). Consequently, the American Astronomical Society (AAS) passed a Council Resolution calling for astronomy programs to eliminate or de-emphasize the use of GRE scores in graduate admissions (Council, 2016).

On the other hand, in meta-analyses of broader data sets than that used in the study of successful early-career astronomers, social scientists have found that the GRE does have predictive power for faculty ratings of graduate student performance, research productivity, and even citation count for published student work (Kuncel & Hezlett, 2007; Kuncel, Wee, Serafin, & Hezlett, 2010).  These studies suggest that the GRE is, in fact, a valid measure of some trait that is relevant to success in graduate study. There is thus a tension between analyses finding predictive power in GRE scores and the existence of highly successful scientists with low GRE scores. Due to the salience of this issue for academic departments that are currently revisiting their admissions requirements, it is thus worthwhile to revisit the caveats that must be applied when interpreting such studies.

Here we will explore two issues relevant to analyses of GRE scores and academic performance, in hopes of bringing wider attention to concepts from quantitative social science. The first issue is that success is usually multifaceted--a person who is weak on one measure might nonetheless be likely to succeed due to other areas of strength. Consequently, the existence of highly successful individuals with low GRE scores (or low scores on any other measure of accomplishment, skill, or ability) is not sufficient to establish that the GRE has poor predictive power. A reasonable alternative hypothesis is that multiple variables are required for accurate predictions of performance. Moreover, the relative significance of different variables can be understood in terms of the economic concept of diminishing marginal returns (Krugman & Wells, 2012). It is thus important to collect data on multiple variables and analyze them in a framework that allows for substitutions and tradeoffs between attributes.

The second issue, range restriction (Hunter, Schmidt, & Le, 2006; Sackett, Lievens, Berry, & Landers, 2007; Wiberg & Sundstrom, 2009), is statistical, arising because student performance data inevitably only includes students who are admitted to a program of study, i.e. a systematically selected subset of all students. A student who is admitted to graduate study in spite of low GRE scores likely has some other significant accomplishment or strength to compensate for low scores, while students with high scores may have been admitted in spite of some other weak point in their applications. If two students have different strengths and weaknesses then a comparison between them does not actually test the effects of a single variable. The result will be weakened correlations between students' performance and their observable characteristics. We will illustrate this issue with computer simulations. Our approach is somewhat similar to previous investigators (Roth et al., 2014) who used computer simulations to demonstrate that range restriction can create apparent differences in the predictive power of test scores for different populations, even if the test is an equally valid predictor of performance for all populations. Among our key findings is that both range restriction and improper weighting of admissions criteria can produce spuriously low (or even negative) correlations between admissions variables and students' subsequent performance.

In what follows, we will explore simple two-variable models of performance, and use computer simulations to examine the consequences of range restriction under conditions of different admissions criteria, including admissions based on an accurate model of student performance, admissions based on cutoffs, and admissions based on multiple variables but with inaccurate weighting of the variables. Our principal finding will be that observed correlations between student performance and student characteristics can depend as much on the criteria used to select students as they do on any causal relationship between admissions variables (or what those variables measure) and academic performance.

*A note on value judgments*

Our analysis does not address whether GRE scores "ought" to be used in admissions. Deriving "ought" prescriptions from data-based "is" statements requires a pre-existing framework of assumptions or commitments (Hume, Green, & Grose, 1964; Searle, 1964). Without such a framework, a particular set of facts does not unambiguously prescribe a course of action. For instance, suppose that we had a highly accurate method for predicting graduate students' performance.[2] The statement "Cosima is very likely to be more academically successful than Helena" would <u>not</u> imply "Cosima ought to be favored over Helena in admissions decisions." To say that Cosima "ought" to be favored we would need to accept the value judgement that admission "ought" to be based on an applicant's likely degree of academic success.

---

2 As evidence that we do not have such a method, this author freely concedes that his graduate school performance was noticeably poorer than what might have been expected from application materials. It is likely that certain faculty privately concur.

There are many possible frameworks for "ought" questions in admissions. A public university might have a legislative charge to focus on particular fields; if Helena's research interests more closely align with that mission then it could be appropriate to favor her. Or, it might be charged with developing talent from all of the communities in the state, and if Helena's community in especial need of inclusion then it might be rational to favor her.

An additional problem is that, in practice, the question "Ought a particular variable be used admissions?" cannot be decoupled from comparisons with alternatives. For instance, researchers have found gender bias in the adjectives in recommendation letters (Madera, Hebl, & Martin, 2009; Jessica M Nicklin & Roch, 2008). These biases often appear against a background of inflated recommendation letters (Jessica M. Nicklin & Roch, 2009), making it likely that negative remarks arising from bias will stand out. Descriptions of research projects may be evaluated more favorably if the writer is believed to be a male (Knobloch-Westerwick, Glynn, & Huge, 2013), raising questions about how fairly application essays might be evaluated. Likewise, science faculty evaluating resumes from hypothetical recent college graduates (i.e. a population similar to applicants for graduate study) rated resumes with male names more favorably than comparable resumes bearing female names (Moss-Racusin, Dovidio, Brescoll, Graham, & Handelsman, 2012). Even requiring research experience poses access issues (Bell et al., 2014).

Nonetheless, a decision must be made, and it must be made using some type of metric. Given that many plausible metrics have identifiable problems, choosing between them requires a mix of "is" statements about consequences and value judgments about the relative desirability of different consequences. It is conceivable that a GRE section "is" a useful predictor of performance but that a rational person might nonetheless conclude that they "ought" to not consider it, because of a conflicting value judgment. Here we will not attempt to derive "ought" statements about the GRE from "is" findings about statistics.

## Background

*Tradeoffs and the multi-faceted nature of talent*

Suppose that for each prospective graduate student we can measure two variables, or student characteristics, which we will call *x* and *y*.[3] We will assume that the correlation between *x* and *y* is weak enough that they can be treated as independent. One can, for the illustrative purposes, suppose that *x* is a "traditional" measure of academic ability and/or accomplishment, such as grade point average or GRE score (whether a subject test, a section of the general test, or a composite of multiple test scores), and that *y* measures some distinct attribute. Possible measures to plot on the *y* axis include less "traditionally academic" measures such as relevant work or research experience, or measures of non-cognitive traits (Kyllonen, Walters, & Kaufman, 2005; W. E. Sedlacek, 2004) such as "grit" (Duckworth, Peterson, Matthews, & Kelly, 2007; Powell, 2013)[4] or "growth mindset" (York, Gibson, & Rankin, 2015). We assume that *x* and *y* are correlated with some measure of student performance, i.e. our dependent variable, which we will denote as *P*. *P* can be defined and measured in many different ways (Gardner, 2009; York, et al., 2015); identifying the particular aspect of performance that one measures is not essential for our purposes. Since performance is difficult to predict with perfect accuracy, we will assume that a predicted performance level *P* calculated from measured student characteristics only represents an <u>average</u> level of performance among students with the same *x* and *y* scores.

---

3 Real admissions processes will typically involve more than two types of data, but two variables suffice to illustrate the relevant issues.

4 It is important to note that the pioneer of grit research, Dr. Angela Lee Duckworth, has issued cautionary remarks about efforts to assign stakes to measures of grit (Duckworth, 2016; Powell, 2013). Nonetheless, we mention grit here because critics of the GRE have issued calls to "diminish reliance on GRE and instead augment current admissions practices with proven markers of achievement, such as grit and diligence" (Miller & Stassun, 2014).

Different combinations of *x* and *y* values could give the same average level of performance *P*. A curve of constant *P* in the (*x*, *y*) plane is closely related to the concept of an "indifference curve" in microeconomics (Krugman & Wells, 2012). While the specific form of such a curve is an empirical issue, its basic shape can be understood on the basis of the economic concept of a diminishing marginal rate of substitution between attributes (Krugman & Wells, 2012). A diminishing marginal rate of substitution is, in turn, closely related to the concept of diminishing returns in production. Economic models typically assume that the production of a good (whether a commodity like wheat, a machine like a car, or a more abstract good like a student-authored research article) exhibits diminishing returns on inputs. This amounts to assuming that successful work is multi-faceted: No single factor is sufficient for success, so as one input becomes plentiful other factors become comparatively more important. For instance, adding even more machines will produce minimal improvements in a factory that is short of machine operators, just as adding more people to an under-equipped factory will be of similarly weak benefit. Likewise, replacing a "smart but lazy" researcher with one who is even smarter but still lazy will produce little improvement in a laboratory's output. In each case, improvements in the factor that is most lacking will produce greater effects than augmenting the factor that is already present in abundance.

A curve that exhibits a diminishing marginal rate of substitution of traits is shown in Figure 1(A) for a hypothetical case with research experience and a GRE score as predictors of performance.[5]

---

5 We chose "GRE" as the name for the *x* variable and "research experience" as the name for the *y* variable simple for convenience in illustrating admission on the basis of two distinct traits. The key thing is that each variable is one that might plausibly be considered in admissions decisions, and they are likely to measure distinct traits.

Because the curve is convex (sloping downward and leveling off), successive increases in the level of one trait compensate for progressively smaller decreases in the other trait. This. For comparison, we also show a curve with a constant rate of substitution in Figure 1(B). Note that the shape of the curve <u>does not</u> imply that students who score poorly on a traditional academic measure like the GRE will always have little research experience and vice-versa. We are only assuming that <u>for people with equal levels of overall performance</u>, deficits of one trait can be compensated for with other strengths. Whether or not people actually cluster at opposite ends, or if they instead span the entire range, is an empirical question, not answerable by theory.

*Range Restriction*

Range restriction effects occur when one selects a narrow sample (by some measure) from a much wider population (Hunter, et al., 2006; Sackett, et al., 2007; Wiberg & Sundstrom, 2009). For instance, suppose that one wanted to understand the effect of age on athletic performance, but one only studied traditional-age college students. The subjects would fall within a narrow range of ages, so other variables would determine the variation in subjects' athletic performances. Nonetheless, typical toddlers and typical octogenarians both run more slowly than typical healthy twenty year-olds. Consequently, sampling traditional-age college students would be insufficient to determine if/how age affects athletic performance in most circumstances.

Likewise, faculty cannot observe how every conceivable student will perform in their program. They typically only observe students who meet their admissions criteria, and those students will typically be on the high end of the range of characteristics observable by admissions committees. Differences in

---

However, none of our conclusions about range restriction and tradeoffs between measures depend on the specific names chosen for the variables.

the dependent variable (research performance) among those students will therefore be determined, to a large extent, by traits that were not observable by the admissions committee.  There may be substantial variation in those traits precisely because they were not part of the selection process.

## Methods

For a quantitative illustration, we performed computer simulations of admissions processes[6],

generating students whose measurable characteristics *x* and *y* were modeled as random variables with standard deviation 1 (for convenience) and mean 3 (to make negative values rare).  For illustrative purposes and ease of display, all figures show results from a sample of $6*10^3$ students (approximately the number receiving a Bachelor's degree in physics in the United States in 2009 (Mulvey & Nicholson,

2015)), but correlations reported in Tables 1-3 are from simulations of $10^5$ students (for better

precision), and Table 4 shows results from a simulation of $10^6$ students (because only a small fraction of

simulated students were used to compute correlations). In each case (except when considering post-

selection effects) we calibrated admissions criteria to admit 23% of the students in the sample

(approximating the ratio of US citizens in their first-year of physics PhD study in the United States in

2009 (Mulvey & White, 2014) to the number of US citizens receiving a bachelor's degree in physics that year).  While our simulation results have greater precision than could be attained in empirical studies with typical feasible sample sizes, our goal here is to illustrate inescapable pitfalls that arise even under optimal conditions.

We used numbers relevant to US citizens for two reasons.  First, admissions decisions depend as much on funding availability as student quality.  At state universities that do not waive tuition for graduate assistants, international students will be more expensive to support.  Also, some grants and fellowships only support research by students with US citizenship. Thus, there may be financial reasons to apply a different level of scrutiny to international applicants. Second, while international graduate students certainly contribute to the diversity of a graduate program's environment, many of the diversity concerns that intersect with critiques of the GRE are rooted in social, economic, and educational conditions in the United States.

---

6 The Python code used for simulations is available at http://tiny.cc/edxk8x.

In our simulations we assumed that a student's performance *P(x,y)* was governed by the following simple equation, chosen to ensure diminishing returns on each variable and a diminishing rate of marginal substitution between variables, while being simple to implement:

$$P(x, y) = \sqrt{1 + x \cdot y} + \varepsilon \tag{1}$$

where $\varepsilon$ is a normally-distributed random variable with mean 0 and variance 0.5. The variance of $\varepsilon$ was chosen so that $\varepsilon$ accounts for 50% of the variance of *P* and *x* and *y* each explain 25% of the variance individually, in keeping with a meta-analysis finding that GRE subject scores have a correlation *r*=0.5 ($r^2$=0.25) with faculty ratings of graduate student performance (Kuncel & Hezlett, 2007). A diminishing marginal rate of substitution is ensured because curves of constant *P* are curves of constant $x \cdot y$, while the diminishing returns on each variable are assured by the fact that the second derivatives of *P* with respect to *x* and *y* are both negative.

We simulated admission of students under three different types of admissions rules (described below). For each case we assumed (for simplicity) that admitted students were at the top (as measured by admission criteria) of their US undergraduate cohort, computed each admitted student's performance P, and computed correlations between *P*, *x*, *y*, and a performance predictor $\hat{P}$ (a function of *x* and *y* that is used to determine who is admitted) for admitted students. We also divided admitted students into two equal-sized tiers (based on their observable characteristics *x* and *y* at the time of admission), to reflect the fact that not all programs are equally competitive, so observations within a single institution will primarily be comparing students with similar observable characteristics at the time of admission. Moreover, division into tiers also reflects the fact that the standardized test scores of a program's students are known to be correlated with program prestige (Sweitzer & Volkwein, 2009).

Our assumption that all students exceeding a given threshold (as measured by observable characteristics at the time of admission) become graduate students is admittedly somewhat unrealistic; real student cohorts may have somewhat greater variability than considered in our simulation (partially mitigating range restriction effects). In reality, there will be students with similar markers of preparation who will choose other educational and career paths. Consequently, the real cohort of first-year graduate students is broader than in our simulation. Additionally, not all first-year PhD students with US bachelor's degrees will come directly from undergraduate study; some will first acquire Master's degrees and/or work experience, potentially making them better-prepared for PhD study than would be predicted by their observable characteristics when they completed their bachelor's degrees. However, our basic modeling approach is still relevant to real studies (with appropriate caveats in interpretation), because potential areas of strength (or weakness) that are not captured by the measured variables $x$ and $y$ affect $P$ via the noise term $\varepsilon$.

*Admission based on an accurate model of student performance*

Our first simulation considered a "best-case" admissions rule, where admission committees accurately understand how performance depends on observable student characteristics, i.e. they have read enough well-designed studies to know that (in this hypothetical scenario) curves of constant performance in the (*x,y*) plane are just curves of constant $x \cdot y$ (per Equation 1) . They thus predict an applicant's performance as:

$$\hat{P} = x \cdot y \tag{2}$$

This predictor does not exhibit diminishing returns on the individual measures *x* and *y*, but if the number of students that programs can admit is fixed then relative comparisons of applicants are all that matter, so any monotonic function of $x \cdot y$ will suffice. While it is of course true that no real admissions committee will have such precise information on how observable student traits affect educational outcomes, we will show that even in this highly idealized case the phenomenon of range restriction limits the conclusions that can be drawn from available data. The use of idealized simulations to set limits is somewhat similar to that used by White and Ivie (White & Ivie, 2013), who simulated a completely unbiased (with respect to gender) hiring process to determine the expected distribution of women among physics departments. Our goal here is to see how student performance would be correlated with observable student traits if departments had the best possible selection process.

Students were admitted if their predicted level of performance $\hat{P}$ was 33% above the mean for all graduating seniors; we found from trial and error that this resulted in admitting 23% of the simulated students (as discussed above). In order to sort the admitted students into two equal-sized tiers, those whose predicted level of performance was 61% above the mean of the applicant pool were placed into the upper tier, and the other admitted students were placed into the lower tier.

*Admission based on cutoffs*

Our next set of simulations considered admission of students who exceed cutoffs for both measured variables, i.e. students for whom $x \geq x_{cutoff}$ and $y \geq y_{cutoff}$. As in the previous simulation, besides correlations between observed student characteristics (*x* and *y*) and student performance (*P*), we also want to examine correlations between *P* and the predictor $\hat{P}$ used by the admissions committee. However, admission based on a pair of cutoffs does not, on the surface, resemble the use of a single function as in the case of using $\hat{P} = x \cdot y$ in the previous section. Nonetheless, a rule that

$x \geq x_{cutoff}$ and $y \geq y_{cutoff}$ is implicitly a rule that $\min\left(\dfrac{x}{x_{cutoff}}, \dfrac{y}{y_{cutoff}}\right) \geq 1$. We can also rewrite this as an inequality with an adjustable parameter on the right side, i.e. a measure of the stringency of our admissions criteria:

$$\min\left(x, x_{cutoff} \cdot \frac{y}{y_{cutoff}}\right) \geq x_{cutoff} \qquad (3)$$

If we think of an admissions process as being about selecting students whose predicted level of performance exceeds a threshold, then the left side of Equation 2 amounts to a performance predictor $\hat{P}$. It is not the most intuitive mathematical expression that one might write for a predictor of performance, but it is nonetheless implied mathematically when using cutoffs.

The fact that *x* and *y* have different cutoffs implies that the two variables can, in some sense, get different weights, i.e. one cutoff might be more stringent than the other, effectively giving that variable more weight than the other. To quantify this idea of the relative weights of the variables, we will define the following ratio to be the weight given to *x* relative to *y*:

$$w \equiv x_{cutoff} \,/\, y_{cutoff} \qquad (4)$$

Our (inaccurate) predictor of performance is thus:

$$\hat{P} = \min\left(x, w \cdot y\right) \qquad (5)$$

While the weight is multiplied by *y* in the equation, a large value of *w* actually implies greater weight given to *x*, because low values of y can still result in admission (i.e. *y* isn't very important) while low values of *x* will not result in admission (i.e. *x* is very important).

We will consider two cases in this set of simulations:

1. *x* and *y* receive equal weight (*w*=1)

2. The cutoff for *x* is twice as stringent as the cutoff for *y* (*w*=2)

These simulations thereby consider admissions processes that combine two flaws: an incorrect model for the shape of curves of constant *P* and inordinate emphasis on one variable (the GRE, in our example).

For the case where *w*=1 (equal weight to both admissions variables), students were admitted if x and y were 0.046 standard deviations above the average. While this may not seem like a stringent admissions criterion, keep in mind that we are assuming that 23% of students go to graduate school. Half of the simulated applicants have above-average values of *x*, half have above-average values of *y*, and since x and y are uncorrelated 25% have above-average values of both. It is thus necessary to only set the admissions requirements slightly above average in order to admit 25% of the simulated students. Of those admitted applicants, they were sorted into the upper tier if they were 0.426 standard deviations above average for each observable characteristic.

For the case where *w*=2 (greater weight attached to *x*, the GRE score), simulated students were admitted if *x* was at least 0.657 standard deviations above the average and *y* was no worse than 1.2 standard deviations below average. Admitted students were sorted into the upper tier if *x* was 1.11 standard deviations above the average and *y* was no worse than 0.974 standard deviations below average. We found by trial and error that these criteria admitted 23% of students and sorted them into two equal-sized tiers.

*Admission based on a linear model*

Our next set of simulations considered admission based on a linear model of student performance (i.e. a model with a constant marginal rate of substitution). In this model, the predictor used by admissions committees is:

$$\hat{P} = w \cdot x + y \tag{6}$$

where *w* is the weight of *x* relative to *y*. Although this predictor has a constant rate of marginal substitution (per Figure 1B) it is nonetheless worth considering for several reasons. First, it is easy to implement, so it might be tempting for an admissions committee to simplify their work thusly. Second, although this predictor does not match the underlying model of performance, we can use these simulations to determine whether the precise form of the predictor matters as much as the relative weights assigned to the observable student characteristics. In our underlying model of student performance (Equation 1) *x* and *y* get equal weight; this linear predictor has an adjustable weight parameter.

For equal weighting of *x* and *y*, students were admitted if their predicted performance $\hat{P}$ was at least 18% greater than the average among simulated students. Admitted students were sorted into the upper tier if their predicted performance $\hat{P}$ was at least 29% greater than the average among simulated students. When x received twice the weight of y, students were admitted if their predicted performance $\hat{P}$ was at least 19% greater than the average among simulated students, and were sorted into the upper tier if their predicted performance $\hat{P}$ was at least 31% greater than the average among simulated students. We found that these admissions criteria led in each case to 23% of selected students being admitted and the division of admitted students into two equal-sized tiers.

*Post-admission sorting of students*

Finally, we conducted a simulation in which <u>all</u> students were admitted to graduate study, and the top students were subsequently identified based on their performance during graduate school rather than their observable traits before graduate school. The study of astronomy postdocs by Levesque et al. would be a relevant example (Levesque, et al., 2015). In this case we have selection on the dependent variable (student performance) rather than independent variables. In order to mimic such studies, we looked at correlations between subsequent student performance and observable traits (at the time of admission) in the top 5%, 2%, and 1% of all students, not just the 23% who would have been admitted in the other simulations.

# Results

*Admission based on an accurate model of student performance*

Figure 3 shows a scatter plot of admitted students in our first simulation, where students are admitted if their expected performance (predicted using Equation 2) exceeds a threshold. Correlations between students' subsequent performance and their observable characteristics at the time of admission are summarized in Table 1. A number of effects are evident. First, the correlation of each independent variable with observed performance is strongest in the overall pool of domestic students (where the range is widest because it includes low-performing students) and weakest in the lower half of admitted students (where the range is narrowest). Correlations are of intermediate strength in the upper half of admitted students because high-performing outliers are present, widening the range, but are stronger in the entire set of accepted students, which has a wider range than either half of the accepted student pool.

This difference between the upper and lower cohorts of accepted students is not an artifact of *P* in Equation 1 having a functional form that exhibits diminishing returns on measured student characteristics. <u>In fact, diminishing returns reduces the variability of performance in the upper half of accepted students more than it reduces the variability in the lower half.</u> Thus, a model with constant or increasing returns on student characteristics would actually exhibit even sharper range restriction effects, and would show even greater contrasts between the upper and lower halves.

It is important to note that in our simulation range restriction is not a restriction on the range of outcomes (i.e. not a restriction on the dependent variable)--even individuals with similar prior accomplishment or preparation can exhibit widely varying levels of performance once admitted (modeled via noise in our simulation)--but rather a restriction on the range of measured characteristics at the time of admission (i.e. a restriction on the independent variables). Because admitted students (both in simulations and in the real world) will have a wide range of post-admission outcomes but a comparatively narrower range of observable characteristics at the time of admission, it can be difficult to observe meaningful correlations between admissions variables and subsequent performance within that admitted pool. However, that does not prove that their measured characteristics at the time of admission had no role in their subsequent outcomes.

Also, while *x* and *y* are uncorrelated in the wider pool of students, range restriction leads to a negative correlation between them in the pools of admitted students. Admitted students who score poorly on one measure are essentially guaranteed to score well on the other, due to the selection process that they went through. Whether traits such as "book smarts" and laboratory skills are negatively correlated among real students is an empirical rather than theoretical question, but the statistical phenomena illustrated here demonstrate that empirical exploration of this question requires a large and carefully-chosen sample to avoid spurious negative correlations. In the absence of careful empirical work, range restriction is likely to reinforce negative narratives about "book smarts" (Mackay, 2015). While such narratives certainly have cultural roots (Hofstadter, 1963) that are not specific to the natural science fields that have seen recent calls for decreased use of GRE scores in admissions, the nature of admission processes ensures that there will be no shortage of anecdotes to reinforce such narratives.

On the other hand, if we examine correlations with the product of GRE score and research experience in Table 1, we see that this product has a stronger correlation with subsequent performance than any other variable. This arises because *P* in Equation 1 explicitly depends upon the product $x \cdot y$ (GRE*Experience) so that product should be more strongly correlated with *P* than either independent variable singly. In real decisions, of course, performance may not be predictable from anything as simple as the product of two numbers, but if admissions are done on some sort of "sliding scale" that admits students who are strong by one measure but weaker by another then there is implicitly a curve with a shape akin to that in Figure 1A, and students above that curve will be admitted. That curve specifies some model for how the admissions committee believes that performance is related to observable characteristics (e.g. grades, experience, test scores, personality and work ethic as described in letters and essays, etc.). If the curve corresponds to an accurate model of academic performance then the correlation between performance and the function specifying that curve should be strong.

*Admission based on cutoffs*

Figure 4 and Table 2 show simulations of admissions based on the performance predictor in Equation 5. As before, correlations are stronger in the full pool of admitted students than in either subset, and stronger in the upper tier than in the lower tier. The correlations between the two admissions variables and subsequent student performance are of similar magnitude when the variables get equal weight (Table 2), but when the GRE ($x$) gets more weight it exhibits a much weaker correlation with performance, while experience (which now gets less weight) exhibits a stronger correlation with performance. The reason for the stronger correlation between experience and performance in the second cutoff model is that there are admitted students who performed poorly in part because of inexperience, but their high GRE scores helped them secure admission; their inexperience and subsequent poor performance are both observable and hence strengthen the observed correlation between experience and performance.

Also, the concave nature of the lower tier means that a spurious negative correlation between GRE score and experience is again found. As before, this negative correlation is unrelated to the underlying sample, and reflects only the selection process. The predictor $\hat{P}$ from Equation 5 does show a positive correlation with observed student performance, but the correlation is never as strong as for $x \cdot y$ (GRE*experience), because this selection process was not designed to use the best available predictor of performance.

*Admission based on a linear model*

Results from simulations of admissions based on a linear model are shown in Figure 5 and Table 3.

Under the linear admissions model with equal weighting, correlations between admissions variables and subsequent performance are similar to those for admissions based on a more accurate performance model.  As before, correlations between observable characteristics and students' subsequent performance are stronger in the full cohort of admitted students than in either subset of admitted students, and are stronger in the upper tier of admitted students than in the lower tier.  Moreover, the shape of the admissions regions in Figure 5A again leads to a negative correlation between GRE score and experience among admitted students, with the negative correlation between strongest in the lower tier and weakest in the full pool of admitted students.  On the other hand, when the GRE receives disproportionate weight it exhibits no significant correlation with student performance in the full cohort and upper tier of admitted students, but shows a substantial negative correlation with performance in the lower tier of admitted students.  This negative correlation arises from the fact that the excessive emphasis placed on GRE scores led to the admission of students with little prior research experience.

Also, in Table 3 the true variable governing performance (product of GRE score and experience) exhibits a stronger correlation with observed student performance than the (inaccurate) linear predictor $\hat{P}$.  However, when the two observed student characteristics each receive their proper weight the difference in correlation is quite small.  The lesson seems to be that getting the precise shape of the admissions curves correct is less important than having curves that satisfy two criteria:

1. The independent variables (i.e. observable student characteristics at the time of admission) receive proper relative weights.

2. One measure can compensate for another, i.e. the curve slopes down and to the right.  Although a decreasing marginal rate of substitution would be ideal, it may be less important than the overall orientation of the curve.

*Post-admission sorting of students*

Table 4 and Figure 6 show the results of a simulation in which all students are admitted to graduate study and the most successful ones are subsequently identified. The samples in Figure 6 are wider than those in previous figures, and so the correlations in Table 4 are indeed stronger than in most previous examples. Nonetheless, correlations between observable characteristics at the time of admission and students' subsequent performance are still suppressed relative to the full pool, while spurious negative correlations between characteristics again appear. This illustrates that selection on the dependent variable can give rise to range restriction effects just as can selection on independent variables.

## Discussion

Our simulations demonstrate that range restriction via selection processes causes correlations between observable student characteristics and subsequent performance depend as much on the selection process as they do on a causal relation between student characteristics and performance. This is consistent with the fact that studies of single departments or programs vary widely in their conclusions about which variables predict which aspects (if any) of graduate school performance. Some studies of individual programs report significant correlations between GRE scores and graduate grades (Burmeister et al., 2014; House, 1999), or between GRE scores and faculty ratings (Burmeister, et al., 2014; Weiner, 2014). Conversely, one can find well-designed studies that show little or no predictive power for the GRE within a single program (Sternberg & Williams, 1997).

Because of the small sample in a study of a single program, it is important to balance the microscopic view afforded by a local study with the big-picture view that can be obtained from meta-analysis using statistical techniques to correct for the effects of range restriction. In a large-scale meta-analysis spanning many disciplines, Kuncel (Kuncel, et al., 2010) found that GRE scores predict performance both in the more coursework-focused MS and the less-structured PhD.  They also found correlations with a wide range of performance measures across a wide range of fields (Kuncel & Hezlett, 2007).  The critical responses to the 2007 meta-analysis, and the rebuttals by Kuncel, are worth reading (Lerdau et al., 2007).  Meta-analyses conducted by a single research group are, of course, not definitive-- a finding should not be trusted without replication by independent investigators.  Nonetheless, it is consistent with the trend found in the simulations above, in which the strongest correlations come from the largest samples.

There are two appropriate responses to range restriction when conducting and interpreting research on the predictive power of the GRE (or any other proposed predictor of student performance). On an interpretive level, one should regard reported correlations with caution, seek information on how students were selected, and look for indications that a study includes efforts to correct for range restriction.  On a technical level, investigators should sample as widely as possible, measure more than one independent variable (if possible), and compare variances within their sample to the wider population from which it is drawn.  There are methods to correct computed correlations for range restriction (Hunter, et al., 2006; Schmidt & Hunter, 2014; Wiberg & Sundstrom, 2009), but the proper use of such methods depends on whether the restriction was applied to the independent variables (e.g. characteristics at the time of admission), dependent variables (e.g. a study that only looks at students who succeeded and graduated, but minimizes selection on the independent variables by including students from less-selective programs) or both (e.g. a study of people who successfully completed highly-selective programs).  Analyses that fail to account for these effects may miss important information.

Perhaps most importantly, it is essential to exmaine more than one independent variable, i.e. more than one student characteristic at the time of matriculation to a graduate program. Comparing people with different GRE scores, in the hopes of determining if there is a measurable relationship between their GRE scores and subsequent performance, is only a meaningful exercise if all other factors are held constant.  If, for instance, students with lower GRE scores are only admitted when they demonstrate strength by some other measure, then all other factors are <u>not</u> held constant.  Conversely, if students with lower GRE scores also tend to suffer from various social or economic disadvantages, a simple comparison of GRE scores and measures of subsequent performance will not suffice to determine if GRE scores measure a cognitive trait that affects performance or other life factors that also affect performance.  In order to make such a determination, one would need to collect data on both GRE scores and also socioeconomic variables, and then compare the performance of people with comparable socioeconomic backgrounds but different GRE scores.

## Conclusions

As stated above, this article does not advance an "ought" position on using standardized tests, nor does it make a definitive "is" statement on the predictive power of standardized tests.  The primary points of this article are:

1.  If performance is determined by multiple variables then the existence of highly successful people with low test scores is insufficient to demonstrate that test scores have little or no predictive power.
2.  Range restriction effects should temper any conclusions drawn from a narrow sample.  Observed correlations between student performance and observable traits will be governed as much by selection processes as by the actual predictive power of the traits in question.  Without carefully examining how study subjects were selected, a low correlation between a particular variable and student performance is insufficient to establish that the variable is unrelated to student performance, while a strong correlation suggests that the variable may be receiving insufficient weight in admissions.

While we have couched our examples in terms of the predictive power of the GRE (as it is a topic of substantial current interest), these caveats also apply to educational questions unrelated to high-stakes admissions. For instance, suppose that one studied the relationship between performance in an introductory science course (taken by students from many different majors) and performance in a more advanced course in the same field (taken mostly by students majoring in the field). Suppose also that most students believe that engineering offers better career prospects than basic science, so the engineering program copes with the flood of aspiring engineers by requiring students to meet certain benchmarks in their first year. Advanced courses in a basic science field will thus include students who were unable to continue in engineering. Moreover, a student who performed well in introductory science but was nonetheless disqualified from engineering may have significant weakness that was not revealed by their grade in introductory science. Such an effect could partially mask a causal relationship between success in the introductory course and success in the more advanced course.

Finally, even if well-conducted studies find that standardized tests have predictive power after accounting for range restriction and the multi-dimensional nature of achievement, one could still rationally refrain from using standardized tests because of a value judgment (W. E. Sedlacek, 2004). However, "is" statements on the predictive power of a test require studies that account for the multivariate nature of talent and success, and that account for statistical issues such as range restriction.

## Notes

The author thanks Homeyra Sadaghiani, Qing Ryan, Kyler Kuehn, and James Hanley for critiquing drafts of this article.

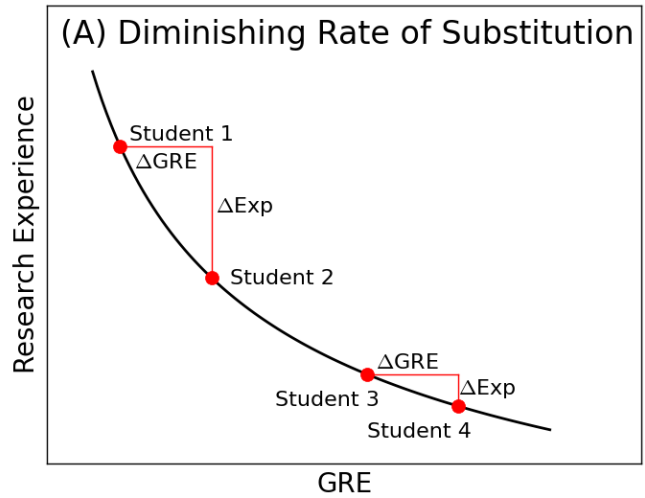

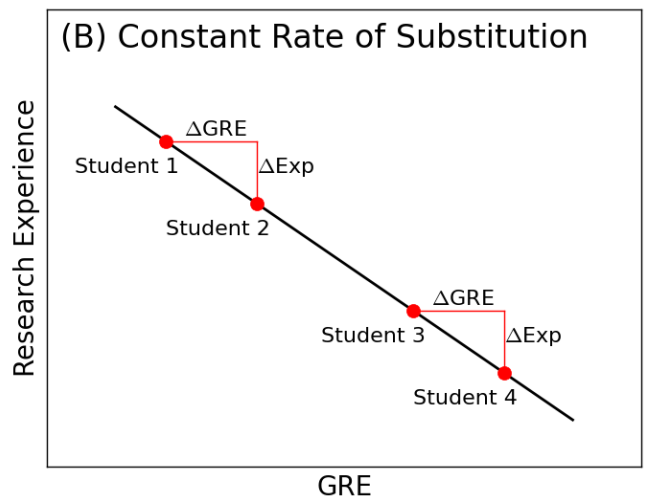

**Figure 1:** Tradeoffs between GRE and research experience as predictors of performance in graduate school. The effects of a constant change in GRE are shown for comparisons between two different pairs of students, for both (A) diminishing marginal rate of substitution and (B) constant marginal rate of substitution.

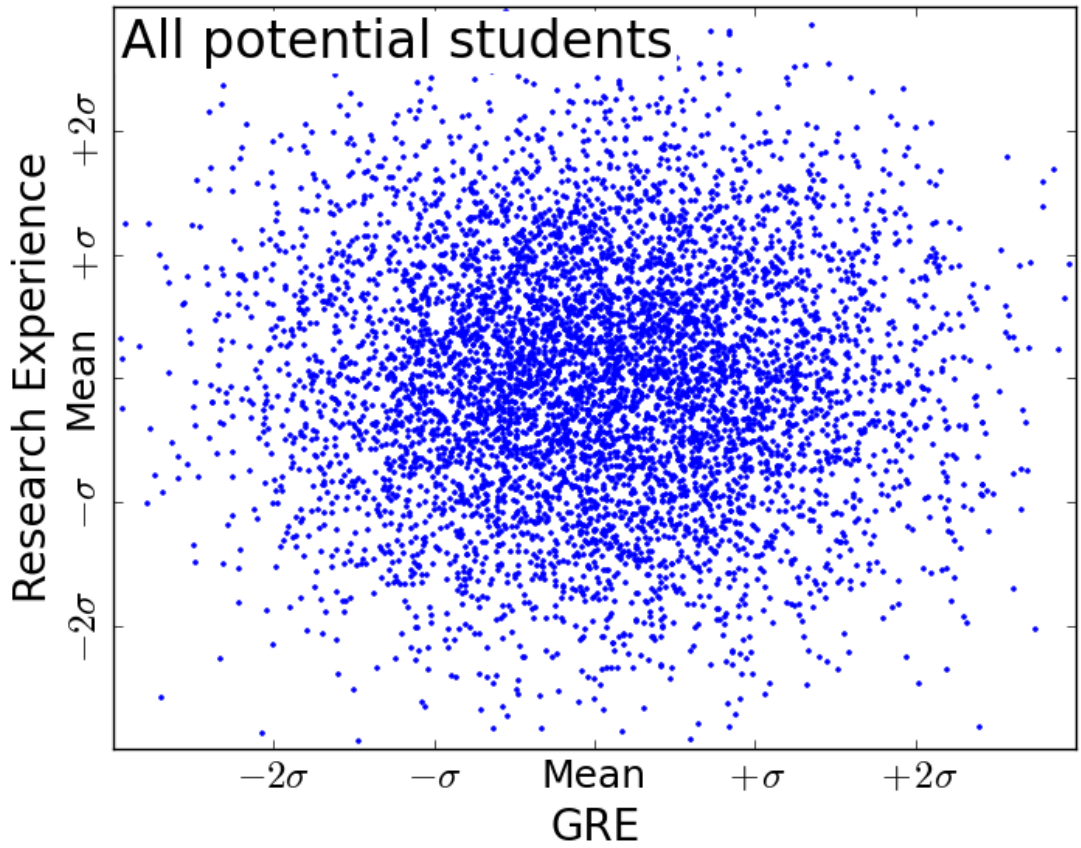

**Figure 2:** The sample used for a simulation of 6,000 students, showing GRE score and research

experience. Both variables are normally distributed, with the same mean and standard deviation.

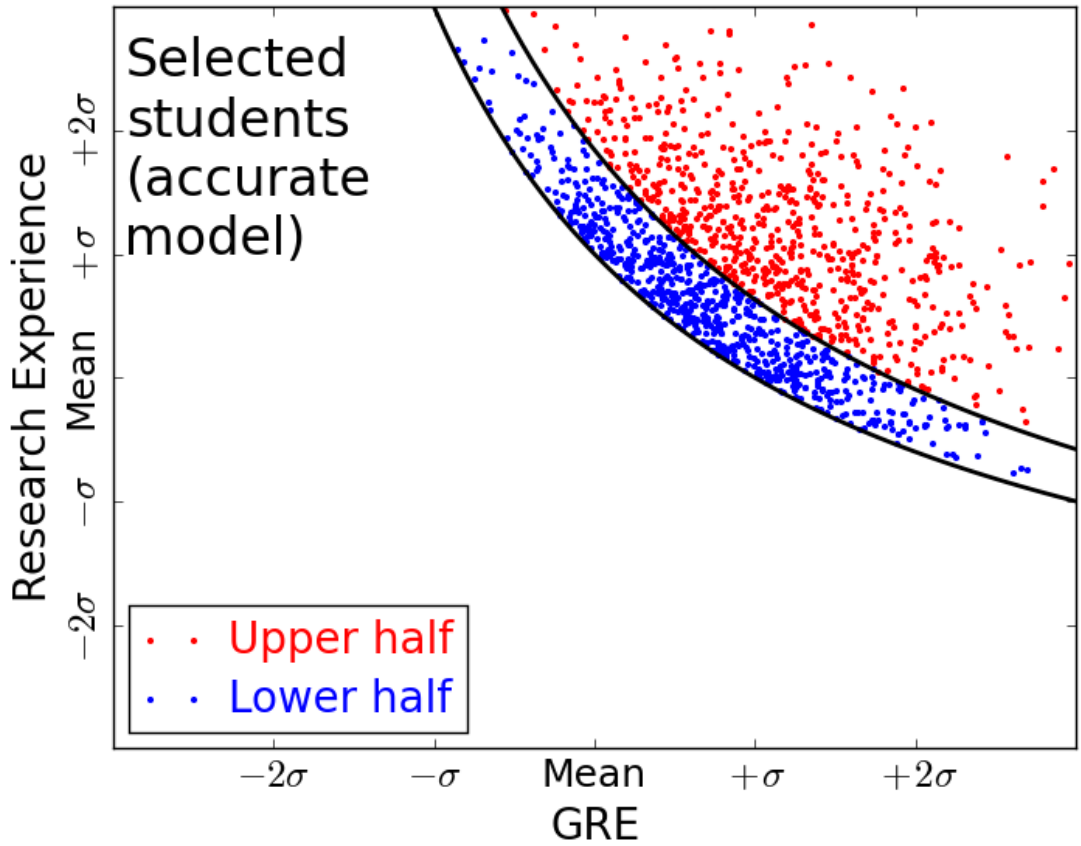

**Figure 3:** The sample from Figure 2 after admission based on an accurate model of student performance

(Equation 2). 23% of the students remain. Admitted students are sorted into two tiers.

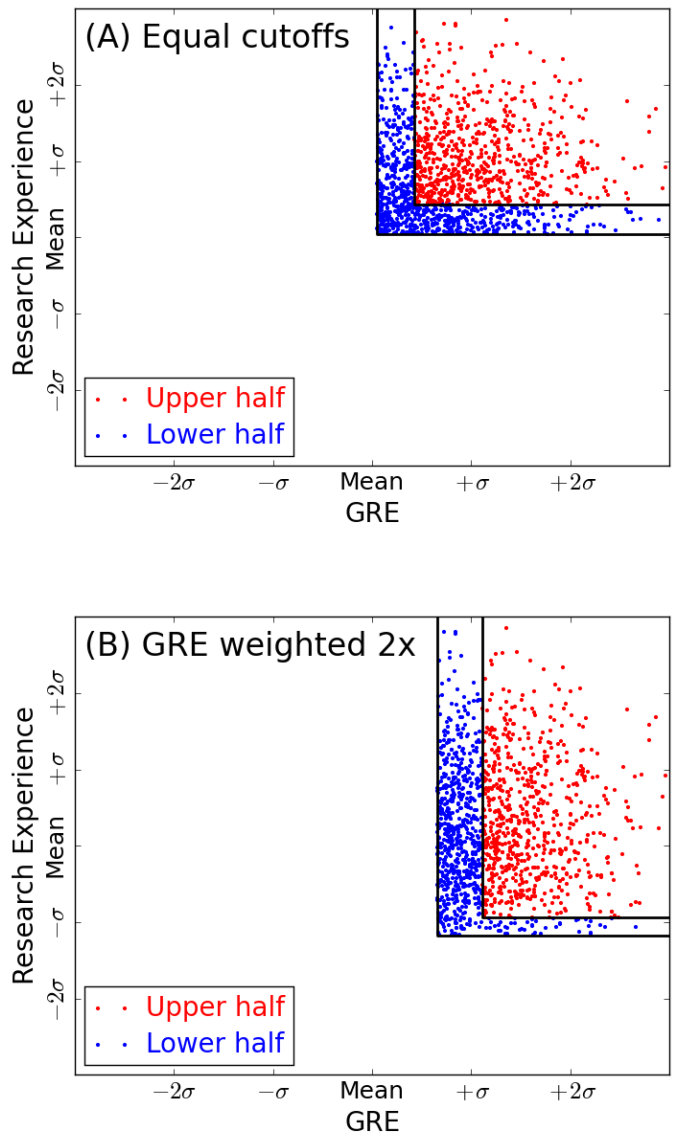

**Figure 4:** The sample from Figure 2 after admissions based on cutoffs for two variables (Equation 5). 23% of the students remain. Admitted students are sorted into two tiers. (A) Equal weight for each variable. (B) GRE given twice the weight of Research Experience.

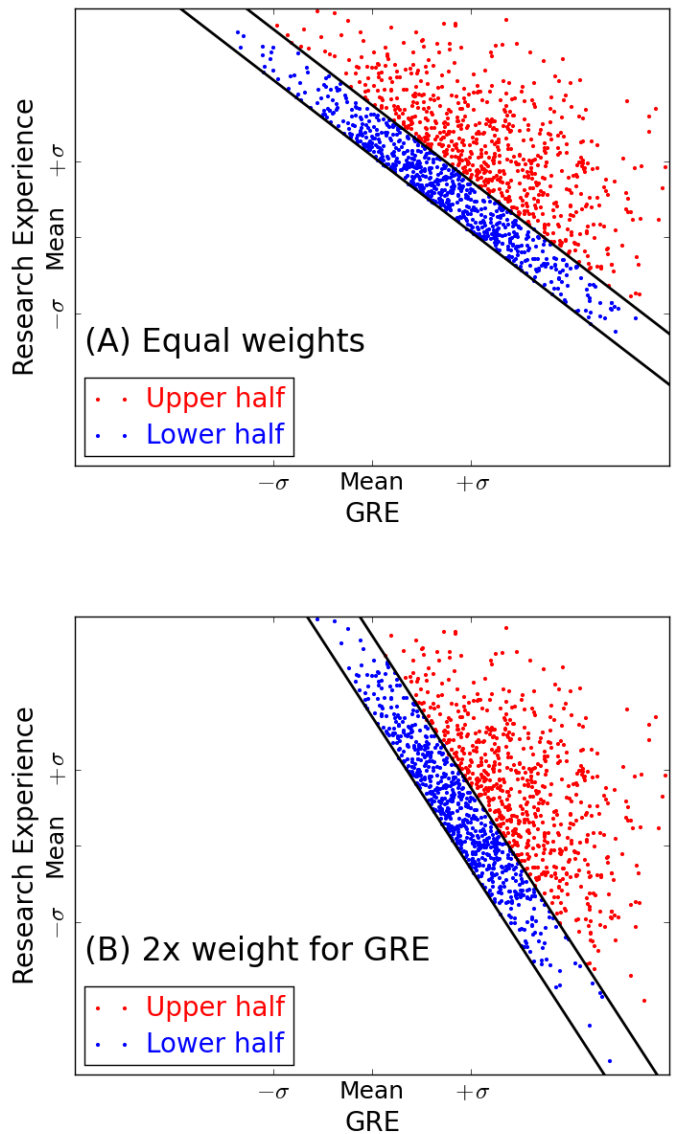

**Figure 5:** The sample from Figure 2 after admissions based on a linear model of student performance (Equation 6). 23% of the students remain. Admitted students are sorted into two tiers. (A) Equal weight for each variable. (B) GRE given twice the weight of Research Experience.

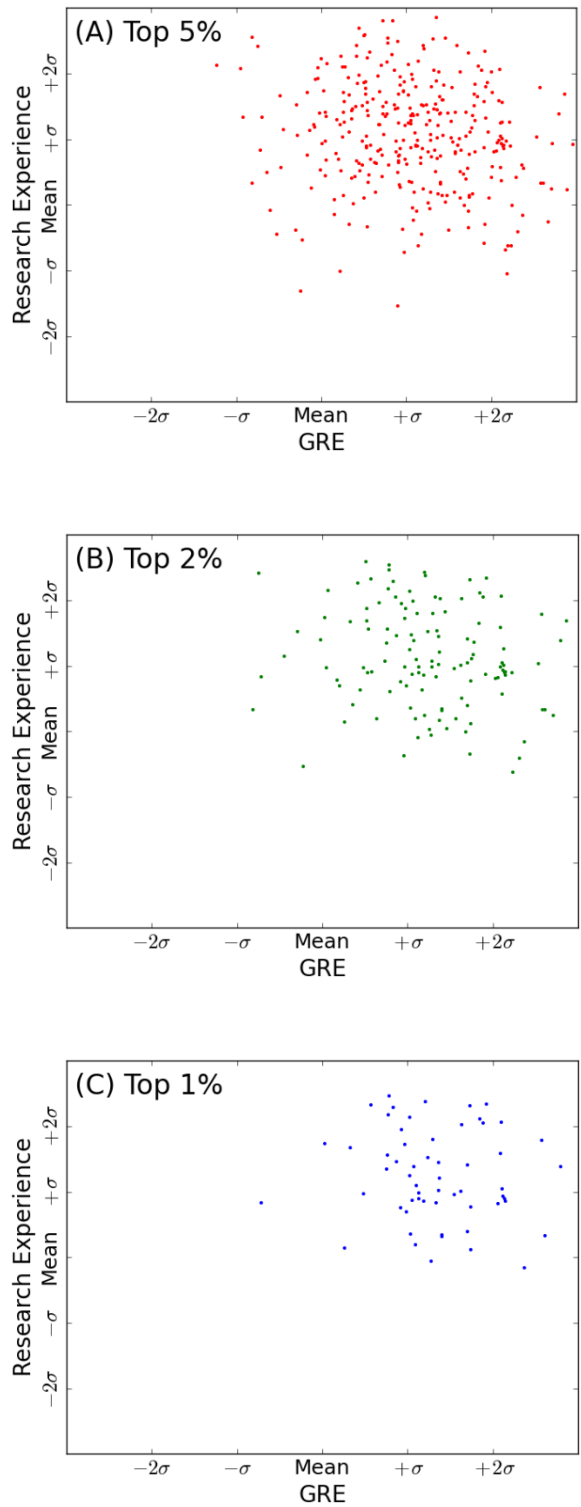

**Figure 6:** The sample from Figure 2 after selecting students with the greatest observed (not predicted) performance. (A) Top 5%, (B) Top 2%, and (C) Top 1%.

TABLE 1

*Simulations of admissions based on an accurate model of student performance*

| Cohort | $P$, GRE | $P$, Exp. | $P, \hat{P}$ | P, GRE*Exp. | GRE, Exp. |
|---|---|---|---|---|---|
| All Students | 0.502 | 0.496 | 0.709 | 0.709 | -0.002 |
| | | | | | |
| Admitted Students | | | | | |
| All | 0.219 | 0.209 | 0.449 | 0.449 | -0.526 |
| Top half | 0.181 | 0.160 | 0.397 | 0.397 | -0.613 |
| Lower half | 0.017 | 0.018 | 0.122 | 0.122 | -0.928 |

Note.—The committee's performance predictor $\hat{P}$ was computed from Equation 2. The number of students used to compute correlations was $10^5$.

Table 2

*Simulations of graduate school admissions based on cutoffs for both variables*

| Cohort | *P*, GRE | *P*, Exp. | $P, \hat{P}$ | P, GRE*Exp | GRE, Exp |
|---|---|---|---|---|---|
| | | | Correlation of | | |
| All Students | 0.502 | 0.496 | | 0.709 | -0.002 |
| | | | | | |
| Admitted Students (equal cutoffs for GRE, Exp.) | | | | | |
| All | 0.364 | 0.345 | 0.416 | 0.503 | -0.003 |
| Top half | 0.335 | 0.320 | 0.373 | 0.467 | -0.015 |
| Lower half | 0.202 | 0.166 | 0.118 | 0.340 | -0.413 |
| | | | | | |
| Admitted Students (2X weight for GRE) | | | | | |
| All | 0.245 | 0.541 | 0.310 | 0.594 | 0.001 |
| Top half | 0.212 | 0.534 | 0.271 | 0.574 | 0.005 |
| Lower half | -0.054 | 0.540 | 0.092 | 0.548 | -0.263 |

Note.—The committee's performance predictor $\hat{P}$ was computed from Equation 5. The number of students used to compute correlations was $10^5$.

TABLE 3

*Simulations of graduate school admissions based on a linear model of student performance*

| Cohort | Correlation of | | | | |
| | $P$, GRE | $P$, Exp. | $P, \hat{P}$ | P, GRE*Exp | GRE, Exp |
|---|---|---|---|---|---|
| All Students | 0.502 | 0.496 | | 0.709 | -0.002 |
| | | | | | |
| Admitted Students (equal weight for GRE, Exp.) | | | | | |
| All | 0.197 | 0.190 | 0.446 | 0.462 | -0.624 |
| Top half | 0.157 | 0.150 | 0.398 | 0.414 | -0.701 |
| Lower half | 0.020 | 0.011 | 0.119 | 0.185 | -0.965 |
| | | | | | |
| Admitted Students (2X weight for GRE) | | | | | |
| All | 0.027 | 0.427 | 0.396 | 0.536 | -0.537 |
| Top half | -0.069 | 0.432 | 0.338 | 0.516 | -0.624 |
| Lower half | -0.301 | 0.359 | 0.136 | 0.394 | -0.945 |

Note.—The committee's performance predictor $\hat{P}$ was computed from Equation 6. The number of students used to compute correlations was $10^5$.

TABLE 4

*Simulation with admission of all students and subsequent identification of the most successful*

| Cohort | Correlation of | | | |
|---|---|---|---|---|
| | *P*, GRE | *P*, Exp. | P, GRE*Exp. | GRE, Exp. |
| All Students | 0.501 | 0.501 | 0.711 | -0.001 |
| Top 5% | 0.238 | 0.229 | 0.370 | -0.148 |
| Top 2% | 0.219 | 0.201 | 0.332 | -0.150 |
| Top 1% | 0.201 | 0.188 | 0.309 | -0.155 |

Note.—Simulations used a pool of $10^6$ students for precision.